\documentclass[]{article}
\usepackage{nips10submit_e}
\nipsfinalcopy

\usepackage{amsmath,amssymb}
\usepackage[numbers]{natbib}
\newcommand{\citeg}[1]{\citep[e.g.][]{#1}}
\usepackage{bibhacks}
\usepackage{nicefrac}
\usepackage{url}
\usepackage{color}
\usepackage{graphicx}
\usepackage{booktabs}
\usepackage{algorithm}
\usepackage[noend]{algpseudocode}
\usepackage{subfig}

\renewcommand{\algorithmicensure}{\textbf{Input:} }
\renewcommand{\algorithmicrequire}{\textbf{Output:} }
\renewcommand\algorithmicthen{}

\newcommand{\gv}{\mathbf{f}}

\newcommand{\gvv}{\mathbf{g}}

\newcommand{\bx}{\mathbf{x}}

\newcommand{\bmm}{\mathbf{m}}

\newcommand{\etab}{\boldsymbol{\eta}}
\newcommand{\bnu}{\boldsymbol{\nu}}
\newcommand{\bomega}{\boldsymbol{\omega}}

\newcommand{\D}{\mathrm{data}}

\newcommand{\N}{\mathcal{N}}

\newcommand{\U}{\mathrm{Uniform}}

\newcommand{\ave}[2]{{\mathbb E}_{#1}\kern-2pt\left[ #2 \right]}
\newcommand{\oo}[1]{\frac{1}{#1}}
\newcommand{\eye}{{\mathbb I}}

\newcommand{\lh}{\mathcal{L}}

\newcommand{\te}{\!=\!}
\newcommand{\tp}{\!+\!}
\newcommand{\tm}{\!-\!}
\newcommand{\ttimes}{\!\times\!}

\newcommand{\la}{\!\leftarrow\!}
\newcommand{\tsim}{\!\sim\!}

\newcommand{\trightarrow}{\!\rightarrow\!}

\newcommand{\g}{\kern1pt|\kern1pt}

\newcommand{\Sprior}{\Sigma_\theta}
\newcommand{\Lprior}{L_{\kern-1pt\Sprior}}
\newcommand{\Spriorp}{\Sigma_{\theta'}}
\newcommand{\Lpriorp}{L_{\kern-1pt\Spriorp}}
\newcommand{\Sppost}{R_\theta}
\newcommand{\Lppost}{L_{\kern-1pt\Sppost}}
\newcommand{\Sppostp}{R_{\theta'}}
\newcommand{\Lppostp}{L_{\kern-1pt\Sppostp}}
\newcommand{\mppost}{\bmm_{\theta\kern-1pt,\gvv}}
\newcommand{\Snoise}{S_\theta}
\newcommand{\Snoisep}{S_{\theta'}}
\newcommand{\ph}{p_h\kern-1pt}

\DeclareMathOperator*{\argmax}{argmax}

\newcommand{\posterior}{p^\star\kern-1pt}

\title{Slice sampling covariance hyperparameters\\ of latent Gaussian models}

\author{ {\bf Iain Murray} \\
School of Informatics\\
University of Edinburgh
\And
{\bf Ryan Prescott Adams} \\
Dept.\ Computer Science\\
University of Toronto
}

\begin{document}

\maketitle

\begin{abstract}
  The Gaussian process (GP) is a popular way to specify dependencies
  between random variables in a probabilistic model. In the Bayesian
  framework the covariance structure can be specified using unknown
  hyperparameters. Integrating over these hyperparameters considers
  different possible explanations for the data when making
  predictions. This integration is often performed using Markov chain
  Monte Carlo (MCMC) sampling. However, with non-Gaussian observations
  standard hyperparameter sampling approaches require careful
  tuning and may converge slowly.  In this paper we
  present a slice sampling approach that requires little tuning while
  mixing well in both strong- and weak-data regimes.
\end{abstract}

\section{Introduction}

Many probabilistic models incorporate multivariate Gaussian
distributions to explain dependencies between variables.  Gaussian
process (GP) models and generalized linear mixed models are common
examples. For non-Gaussian observation models, inferring the
parameters that specify the covariance structure can be
difficult. Existing computational methods can be split into
two complementary classes: deterministic approximations and Monte
Carlo simulation. This work presents a method to make the sampling
approach easier to apply.

In recent work
\citet{murray2010} developed a slice sampling \citep{neal2003a} variant,
\emph{elliptical slice sampling}, for updating strongly
coupled a-priori Gaussian variates given non-Gaussian
observations. Previously, \citet{agarwal2005} demonstrated the utility
of slice sampling for updating covariance parameters, conventionally
called \textit{hyperparameters}, with a Gaussian
observation model, and questioned the possibility of slice sampling in
more general settings.
In this work we develop a new slice sampler for updating covariance
hyperparameters. Our method uses a robust representation that should work
well on a wide variety of problems, has very few technical requirements,
little need for tuning and so should be easy to apply.

\subsection{Latent Gaussian models}

We consider generative models of data that depend on a vector of
latent variables~$\gv$ that are Gaussian distributed with covariance
$\Sprior$ set by unknown hyperparameters~$\theta$. These models are
common in the machine learning Gaussian process literature
\citeg{rasmussen2005a} and throughout the statistical sciences.  We
use standard notation for a Gaussian distribution with mean~$\bmm$ and
covariance~$\Sigma$,
\begin{equation}
    \N(\gv; \bmm,\Sigma) \equiv |2\pi\Sigma|^{-\nicefrac{1}{2}}
    \exp\kern-1pt\big(
        \vcenter{\hbox{\scalebox{0.8}{$
            -{\textstyle\frac{1}{2}} (\gv\tm\bmm)^\top \Sigma^{-1} (\gv\tm\bmm)
        $}}}
    \big),
    \label{eqn:gaussian}
\end{equation}\\[-0.15in]
and use ${\gv \sim \N(\bmm,\Sigma)}$ to indicate that $\gv$ is drawn
from a distribution with the density in~\eqref{eqn:gaussian}.

The generic form of the generative models we consider is summarized by
\begin{equation*}
\begin{split}
    \text{covariance hyperparameters}
    &~~
    \theta \sim \ph,\\
    \text{latent variables}
    &~~
    \gv \sim \N(0,\Sprior),\\
    \text{and a conditional likelihood}
    &~~
    P(\D\g\gv) = \lh(\gv).
\end{split}
\end{equation*}
The methods discussed in this paper apply to covariances~$\Sprior$
that are arbitrary positive definite functions parameterized
by~$\theta$.  However, our experiments focus on the popular case where
the covariance is associated with $N$~input vectors
$\{\bx_n\}_{n=1}^N$ through the squared-exponential kernel,
\vspace*{-0.1cm}
\begin{equation}
    ({\Sprior})_{ij}
    \kern-1pt=\kern-1pt
    k(\bx_i, \bx_j)
    \kern-1pt=\kern-1pt
    \sigma_f^2 \exp\!\Big(\kern-2pt-
    \kern-1pt
    {\textstyle\oo{2}}
    \kern-1pt
    \textstyle
    \sum_{d=1}^D \kern-3pt
    \frac{(x_{d,i} - x_{d,j})^2 }{ \ell_d^2}
    \kern-1pt
    \Big),
    \label{eqn:se_kernel}
\end{equation}
with hyperparameters $\theta\te\{\sigma_f^2,\{\ell_d\}\}$.
Here $\sigma_f^2$ is the `signal variance' controlling the overall
scale of the latent variables~$\gv$. The $\ell_d$ give characteristic
lengthscales for converting the distances between inputs into
covariances between the corresponding latent values~$\gv$.

For non-Gaussian likelihoods we wish to sample from the joint
posterior over unknowns,
\begin{equation}
    P(\gv,\theta \g \D) =
        \textstyle\frac{1}{Z}\,
    \lh(\gv)\,\N(\gv; 0,\Sprior)\,\ph(\theta)\,.
    \label{eqn:joint_posterior}
\end{equation}
We would like to avoid implementing new code or tuning algorithms for
different covariances $\Sprior$ and conditional likelihood
functions~$\lh(\gv)$.

\vspace*{-0.1cm}
\section{Markov chain inference}
\label{sec:mcmc}
\vspace*{-0.2cm}

A Markov chain transition operator $T(z'\la z)$ defines a conditional
distribution on a new position~$z'$ given an initial position~$z$. The
operator is said to leave a target distribution~$\pi$
invariant if
$\pi(z') \te \int T(z'\la z)\,\pi(z)\;\mathrm{d}z$.
A standard way to sample from the joint
posterior~\eqref{eqn:joint_posterior} is to alternately simulate
transition operators that leave its conditionals,
$P(\gv\g\D,\theta)$ and~$P(\theta\g\gv)$, invariant. Under
fairly mild conditions the Markov chain will equilibrate towards
the target distribution~\citeg{tierney1994a}.

Recent work has focused on transition operators for updating the latent
variables~$\gv$ given data and a fixed covariance~$\Sprior$
\citep{titsias2009,murray2010}.
Updates to the hyperparameters for fixed latent variables~$\gv$ need
to leave the conditional
posterior,
\begin{equation}
  P(\theta\g\gv) \propto \N(\gv; 0,\Sprior)\,\ph(\theta),
\end{equation}
invariant. The simplest algorithm for this is the Metropolis--Hastings
operator, see Algorithm~\ref{alg:fixing-gv}. Other possibilities
include slice sampling \citep{neal2003a} and Hamiltonian Monte Carlo
\citep{duane1987,neal2011}.

\begin{figure}[t]
  \vspace*{-0.5cm}
  \hspace*{-0.01\linewidth}
  \subfloat[Prior draws]{%
  \includegraphics[width=0.5\linewidth]{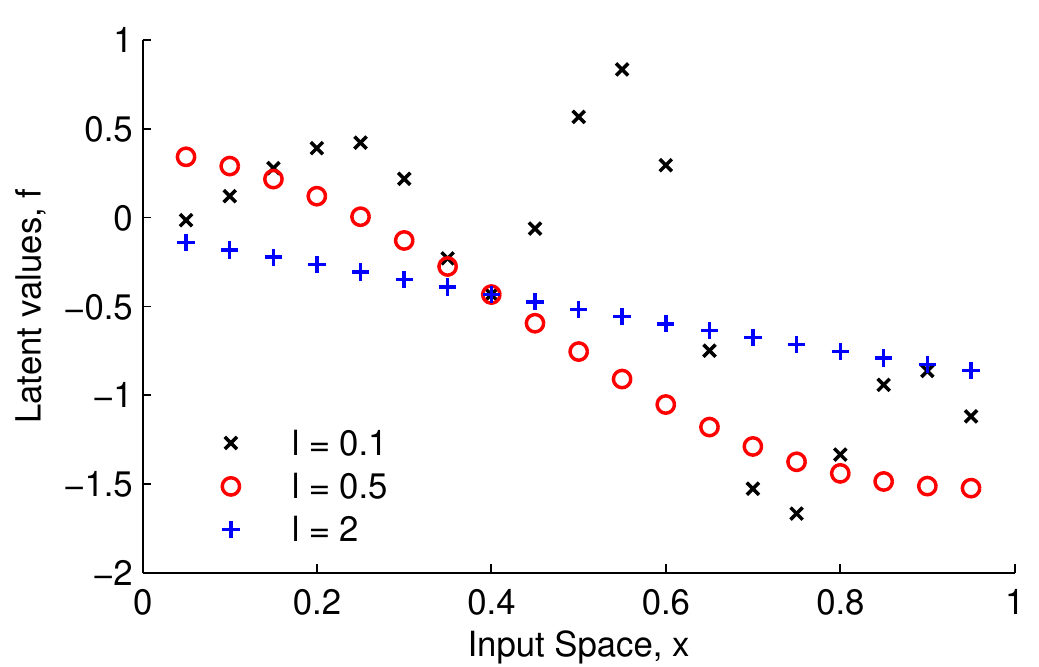}
    \label{fig:priordraws}%
  }~%
  \hspace*{-0.01\linewidth}
  \subfloat[Lengthscale given $\gv$]{%
  \includegraphics[width=0.5\linewidth]{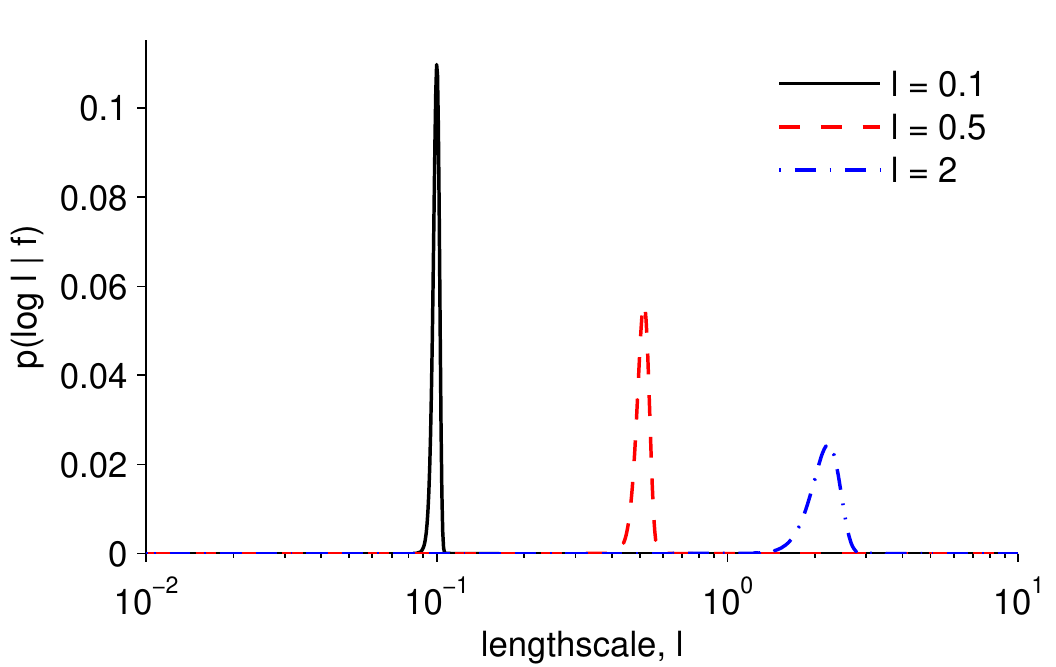}
    \label{fig:postcondf}%
  }
  \vspace*{-0.1cm}
  \caption{\small \textbf{\subref{fig:priordraws}}~Shows draws from
  the prior over~$\gv$ using three different lengthscales in the
  squared exponential covariance~\eqref{eqn:se_kernel}.
  \textbf{\subref{fig:postcondf}}~Shows the posteriors over
  log-lengthscale for these three draws.}
\label{fig:samplingprior}%
\vspace*{-0.1cm}
\end{figure}

Alternately fixing the unknowns $\gv$ and~$\theta$ is appealing from
an implementation standpoint. However, the resulting Markov chain can
be very slow in exploring the joint posterior distribution.
Figure~\ref{fig:priordraws} shows latent vector samples using
squared-exponential covariances with different lengthscales. These
samples are highly informative about the lengthscale hyperparameter
that was used, especially for short lengthscales. The sharpness of
$P(\theta\g\gv)$, Figure~\ref{fig:postcondf}, dramatically limits the
amount that \emph{any} Markov chain can update the
hyperparameters~$\theta$ for fixed latent values~$\gv$.

\subsection{Whitening the prior}
\label{sec:whiten}

Often the conditional likelihood is quite weak; this is why strong
prior smoothing assumptions are often introduced in latent Gaussian
models. In the extreme limit in which there is no data, i.e.\ $\lh$ is
constant, the target distribution is the prior model,
$P(\gv,\theta)\te\N(\gv;0,\Sprior)\,\ph(\theta)$. Sampling from the
prior should be easy, but alternately fixing $\gv$ and~$\theta$ does
not work well because they are strongly coupled.  One strategy is to
reparameterize the model so that the unknown variables are independent
under the prior.

Independent random variables can be identified from a commonly-used
generative procedure for the multivariate Gaussian distribution. A
vector of independent normals, $\bnu$, is drawn independently of the
hyperparameters and then deterministically transformed:
\begin{equation}
  \bnu \sim \N(0, \eye), \qquad
  \gv = \Lprior \bnu, \qquad
  \text{where~} {\Lprior^{\vphantom{\top}}\Lprior^\top\te\Sprior}.
  \label{eqn:sample_prior}
\end{equation}
\textbf{Notation:} Throughout this paper $L_C$ will be any user-chosen
square root of covariance matrix~$C$. While any matrix square root can
be used, the lower-diagonal Cholesky decomposition is often the most
convenient. We would reserve $C^{1/2}$ for the principal square root,
because other square roots do not behave like powers: for example,
$\mathrm{chol}(C)^{-1}\neq \mathrm{chol}(C^{-1})$.

We can choose to update the hyperparameters~$\theta$ for fixed~$\bnu$
instead of fixed~$\gv$. As the original latent variables~$\gv$ are
deterministically linked to the hyperparameters~$\theta$
in~\eqref{eqn:sample_prior}, these updates will actually change
both~$\theta$ and~$\gv$. The samples in Figure~\ref{fig:priordraws}
resulted from using the same whitened variable~$\bnu$ with different
hyperparameters. They follow the same general trend, but vary over the
lengthscales used to construct them.

The posterior over hyperparameters for fixed $\bnu$ is apparent by
applying Bayes rule to the generative procedure
in~\eqref{eqn:sample_prior}, or one can laboriously obtain it by
changing variables in~\eqref{eqn:joint_posterior}:
\begin{equation}
    P(\theta\g \bnu, \D) \propto
    P(\theta, \bnu, \D) =
    P(\theta, \gv\te \Lprior\bnu, \D) \left| \Lprior \right|
    \propto \dots
    \propto \lh(\gv(\theta,\bnu))\,\ph(\theta).
\end{equation}
Algorithm~\ref{alg:fixing-bnu} is the Metropolis--Hastings operator
for this distribution. The acceptance rule now depends on the latent
variables through the conditional likelihood~$\lh(\gv)$ instead of the
prior~$\N(\gv;0,\Sprior)$ and these variables are automatically
updated to respect the prior.  In the no-data limit,
new hyperparameters proposed from the prior are always accepted.

\begin{figure*}[ttt!]
\vspace*{-0.4cm}
\begin{minipage}{\linewidth}%
\scalebox{0.85}{
\begin{minipage}[t]{0.56\linewidth}%
\begin{algorithm}[H]
  \caption{M--H transition for fixed
  $\gv$}%
  \begin{algorithmic}[1]
    \Ensure Current $\gv$ and hyperparameters~$\theta$;\newline proposal dist.~$q$; covariance function $\Sigma_{()}$.
    \Require Next hyperparameters
    \State Propose:
    $\theta' \sim q(\theta'\,;\,\theta)$
    \State Draw $u \sim \U(0,1)$
    \If{${u < \frac{\N(\gv;0,\Spriorp)\,\ph(\theta')\,q(\theta\,;\,\theta')}{
        \N(\gv;0,\Sprior)\,\ph(\theta)\,q(\theta';\,\theta)}}$}
    \State \Return $\theta'$
    \Comment{Accept new state}
    \Else
    \State \Return $\theta$
    \Comment{Keep current state}
    \EndIf
  \end{algorithmic}
  \label{alg:fixing-gv}%
\end{algorithm}%
\end{minipage}
}
\hspace*\fill
\scalebox{0.85}{
\begin{minipage}[t]{0.56\linewidth}
\begin{algorithm}[H]
  \caption{M--H transition for fixed~$\bnu$}%
  \begin{algorithmic}[1]
      \Ensure Current state $\theta$, $\gv$; proposal dist.\ $q$;\newline covariance function $\Sigma_{()}$; likelihood $\lh()$.

    \Require Next $\theta$, $\gv$
    \State Solve for $\N(0,\eye)$ variate: $\bnu\te \Lprior^{-1}\gv$
    \State Propose $\theta' \sim q(\theta'\,;\,\theta)$
    \State Compute implied values: $\gv' \te \Lpriorp\bnu$
    \State Draw $u \sim \U(0,1)$
    \If{${u < \frac{\lh(\gv')\,\ph(\theta')\,q(\theta\,;\,\theta')}{
        \lh(\gv)\,\ph(\theta)\,q(\theta';\,\theta)}}$}
    \State \Return $\theta'$, $\gv'$ \Comment{Accept new state}
    \Else
    \State \Return $\theta$, $\gv$ \Comment{Keep current state}
    \EndIf
  \end{algorithmic}
  \label{alg:fixing-bnu}%
\end{algorithm}
\end{minipage}
}
\end{minipage}
\vspace*{-0.45cm}
\end{figure*}

\section{Surrogate data model}
\label{sec:surrogate}

\begin{figure}[t]
  \centerline{\includegraphics[width=0.45\linewidth]{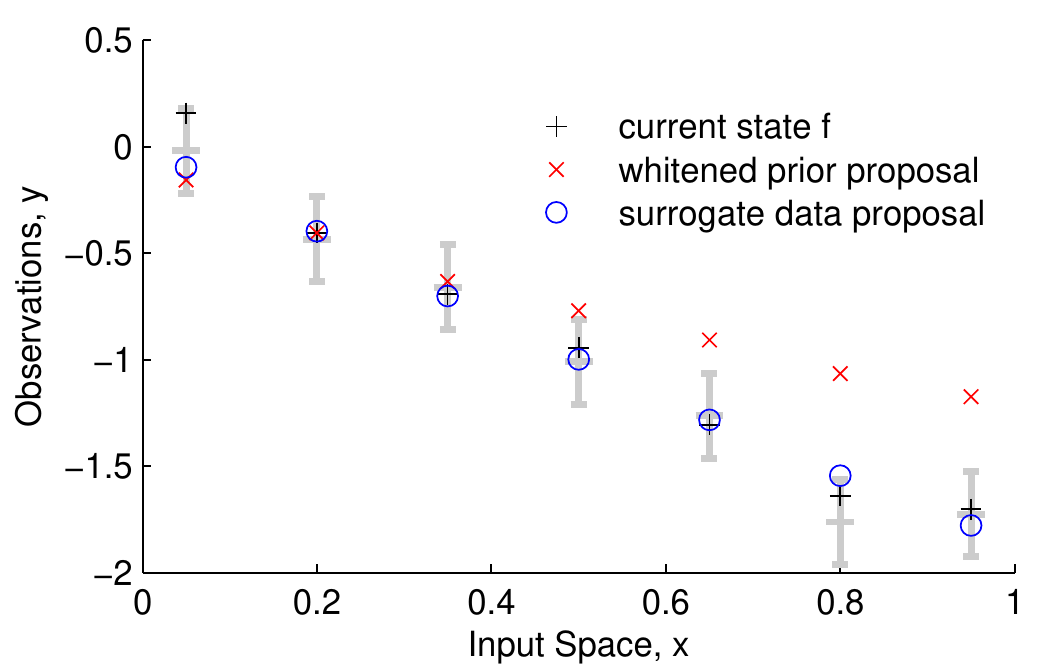}}
  \caption{\small A regression problem with Gaussian observations illustrated
  by $2\sigma$ gray bars. The current state of the sampler has a short
  lengthscale hyperparameter ($\ell\te0.3$); a longer lengthscale
  ($\ell\te1.5$) is being proposed. The current latent variables do not lie on
  a straight enough line for the long lengthscale to be plausible.
  Whitening the prior (Section~\ref{sec:whiten}) updates the latent
  variables to a straighter line, but ignores the observations. A
  proposal using surrogate data (Section~\ref{sec:surrogate}, with
  $\Snoise$ set to the observation noise) sets the
  latent variables to a draw that is plausible for the proposed
  lengthscale while being close to the current state.
  }
\label{fig:tube}%
\end{figure}

Neither of the previous two algorithms are ideal for statistical
applications, which is illustrated in Figure~\ref{fig:tube}.
Algorithm~\ref{alg:fixing-bnu} is ideal in the ``weak data'' limit
where the latent variables~$\gv$ are distributed according to the
prior. In the example, the likelihoods are too restrictive for
Algorithm~\ref{alg:fixing-bnu}'s proposal to be acceptable. In the
``strong data'' limit, where the latent variables~$\gv$ are fixed by
the likelihood~$\lh$, Algorithm~\ref{alg:fixing-gv} would be ideal.
However, the likelihood terms in the example are not so strong that
the prior can be ignored.

For regression problems with Gaussian noise the latent variables can
be marginalised out analytically, allowing hyperparameters to be
accepted or rejected according to their marginal posterior
$P(\theta\g\D)$. If latent variables are required they can be sampled
directly from the conditional posterior $P(\gv\g\theta,\D)$. To build
a method that applies to non-Gaussian likelihoods, we create an
auxiliary variable model that introduces surrogate Gaussian
observations that will guide joint proposals of the hyperparameters
and latent variables.

We augment the latent Gaussian model with auxiliary variables, $\gvv$,
a noisy version of the true latent variables:
\begin{equation}
    P(\gvv \g \gv,\theta) = \N(\gvv;\, \gv, \Snoise).
    \label{eqn:drawg}
\end{equation}
For now $\Snoise$ is an arbitrary free parameter that could be set by
hand to either a fixed value or a value that depends on the current
hyperparameters~$\theta$. We will discuss how to automatically set
the auxiliary noise covariance~$\Snoise$ in
Section~\ref{sec:Snoise}.

The original model, $\gv\sim\N(0,\Sprior)$ and~\eqref{eqn:drawg}
define a joint auxiliary distribution $P(\gv,\gvv\g\theta)$
given the hyperparameters.  It is possible to sample from this
distribution in the opposite order, by first drawing the auxiliary
values from their marginal distribution
\begin{equation}
    P(\gvv\g\theta) = \N(\gvv;\, 0, \Sprior \tp \Snoise),
    \label{eqn:gmarg}
\end{equation}
and then sampling the model's latent values conditioned on the auxiliary
values from
\begin{align}
    P(\gv \g \gvv,\theta) &= \N(\gv;\, \mppost, \Sppost), \text{where some standard manipulations give:}\notag\\
  \Sppost &= (\Sprior^{-1} \tp \Snoise^{-1})^{-1}
  = \Sprior \tm \Sprior(\Sprior \tp \Snoise)^{-1}\Sprior
  = \Snoise \tm \Snoise(\Snoise \tp \Sprior)^{-1}\Snoise, \notag\\
    \mppost &= \Sprior(\Sprior \tp \Snoise)^{-1}\gvv
    = \Sppost^{\vphantom{-1}}\Snoise^{-1}\gvv.
    \label{eqn:pseudopost}
\end{align}
That is, under the auxiliary model the latent variables of interest
are drawn from their posterior given the \emph{surrogate data}~$\gvv$.
Again we can describe the sampling process via a draw from a spherical
Gaussian:
\begin{equation}
  \etab \sim \N(0, \eye), \qquad
  \gv = \Lppost \etab + \mppost, \qquad
  \text{where~} {\Lppost}^{\vphantom{\top}}\Lppost^\top\te\Sppost.
  \label{eqn:sample_ppost}
\end{equation}
We then condition on the ``whitened'' variables~$\etab$ and the
surrogate data~$\gvv$ while updating the hyperparameters~$\theta$. The
implied latent variables $\gv(\theta,\etab,\gvv)$
will remain a plausible draw from the surrogate posterior for the
current hyperparameters. This is illustrated in Figure~\ref{fig:tube}.

We can leave the joint distribution~\eqref{eqn:joint_posterior}
invariant by updating the following conditional distribution derived
from the above generative model:
\begin{equation}
    P(\theta \g \etab, \gvv, \D) \propto
    P(\theta , \etab, \gvv, \D) \propto
     \lh\big(\gv(\theta, \etab, \gvv)\big)\,
    \N(\gvv; 0, \Sprior\tp\Snoise)\,
    \ph(\theta).
    \label{eqn:surrogate_score}
\end{equation}
The Metropolis--Hastings Algorithm~\ref{alg:surr-mh} contains a ratio
of these terms in the acceptance rule.

\subsection{Slice sampling}
\label{sec:slice}
\vspace*{-0.1cm}

The Metropolis--Hastings algorithms discussed so far have a proposal
distribution $q(\theta';\theta)$ that must be set and tuned.  The
efficiency of the algorithms depend crucially on careful choice of the
scale~$\sigma$ of the proposal distribution.  Slice sampling
\citep{neal2003a} is a family of adaptive search procedures that are
much more robust to the choice of scale parameter.

Algorithm~\ref{alg:surr-slice} applies one possible slice sampling
algorithm to a scalar hyperparameter~$\theta$ in the surrogate data
model of this section. It has a free parameter~$\sigma$, the scale of
the initial proposal distribution. However, careful tuning of this
parameter is not required. If the initial scale is set to a large
value, such as the width of the prior, then the width of the proposals
will shrink to an acceptable range exponentially quickly.
Stepping-out procedures~\citep{neal2003a} could be used to adapt
initial scales that are too small. We assume that axis-aligned
hyperparameter moves will be effective, although reparameterizations
could improve performance~\citeg{christensen2006}.

\begin{figure*}[ttt!]
\vspace*{-0.4cm}
\begin{minipage}{\linewidth}%
\scalebox{0.85}{
\begin{minipage}[t]{0.56\linewidth}%
\begin{algorithm}[H]
  \caption{Surrogate data M--H}%
  \begin{algorithmic}[1]
    \Ensure $\theta$, $\gv$; prop.\ dist.\ $q$; model of Sec.~\ref{sec:surrogate}.
    \Require Next $\theta$, $\gv$
    \State Draw surrogate data: $\gvv \sim \N(\gv, \Snoise)$
    \State Compute implied latent variates:\newline
           $\etab\te\Lppost^{-1}(\gv-\mppost)$
    \State Propose $\theta' \sim q(\theta'\,;\,\theta)$
    \State Compute function~$\gv'\te \Lppostp\etab + \boldsymbol{m}_{\theta',\gvv}$
    \State Draw $u \sim \U(0,1)$
    \If{${u < \frac{\lh(\gv')\,\N(\gvv; 0, \Spriorp\tp\Snoisep)\,\ph(\theta')\,q(\theta\,;\,\theta')}{
    \lh(\gv)\,\N(\gvv; 0, \Sprior\tp\Snoise)\,\ph(\theta)\,q(\theta'\,;\,\theta)}}$}
    \State \Return $\theta'$, $\gv'$ \Comment{Accept new state}
    \Else
    \State \Return $\theta$, $\gv$ \Comment{Keep current state}
    \EndIf
  \end{algorithmic}
  \label{alg:surr-mh}%
\end{algorithm}
\end{minipage}
}
\hspace*\fill
\scalebox{0.85}{
\begin{minipage}[t]{0.56\linewidth}
\begin{algorithm}[H]
  \caption{Surrogate data slice sampling}%
  \begin{algorithmic}[1]
    \Ensure $\theta$, $\gv$; scale~$\sigma$; model of Sec.~\ref{sec:surrogate}.
    \Require Next~$\gv$,~$\theta$
    \State Draw surrogate data: $\gvv \sim \N(\gv, \Snoise)$
    \State Compute implied latent variates:\newline
           $\etab\te\Lppost^{-1}(\gv-\mppost)$
    \State Randomly center a bracket:\hspace*\fill\linebreak
    $v \sim \U(0,\sigma),\;\; \theta_{\sf{min}} \te \theta\tm v,\;\;
    \theta_{\sf{max}} \te \theta_{\sf{min}}\tp\sigma$
    \State Draw $u \sim \U(0,1)$
    \State Determine threshold:\newline
           $y = u\,\lh(\gv)\,\N(\gvv; 0, \Sprior\tp\Snoise)\,\ph(\theta)$
    \State Draw proposal: $\theta' \sim \U(\theta_{\sf{min}},
    \theta_{\sf{max}})$
    \label{alg:line:restart}
    \State Compute function~$\gv'\te \Lppostp\etab + \boldsymbol{m}_{\theta',\gvv}$
    \If{$\lh(\gv')\,\N(\gvv; 0, \Spriorp\tp\Snoisep)\,\ph(\theta') > y$}
    \State \Return $\gv'$, $\theta'$
    \ElsIf{$\theta' < \theta$}
    \State Shrink bracket minimum: $\theta_{\sf{min}} = \theta'$
    \Else
    \State Shrink bracket maximum: $\theta_{\sf{max}} = \theta'$
    \EndIf
    \State \textbf{goto} \ref{alg:line:restart}
    \vspace*{-0.1cm}
  \end{algorithmic}
  \label{alg:surr-slice}%
\end{algorithm}
\end{minipage}
}
\end{minipage}
\vspace*{-0.3cm}
\end{figure*}

\vspace*{-0.1cm}
\subsection{The auxiliary noise covariance $\Snoise$}
\label{sec:Snoise}
\vspace*{-0.2cm}

The surrogate data~$\gvv$ and noise covariance~$\Snoise$ define a
pseudo-posterior distribution that softly specifies a plausible region
within which the latent variables~$\gv$ are updated. The noise
covariance determines the size of this region. The first two baseline
algorithms of Section~\ref{sec:mcmc} result from limiting cases
of~$\Snoise\te\alpha\eye$: 1)~if $\alpha\te0$ the surrogate data and
the current latent variables are equal and the acceptance ratio
reduces to that of Algorithm~\ref{alg:fixing-gv}. 2)~as
$\alpha\trightarrow\infty$ the observations are uninformative about
the current state and the pseudo-posterior tends to the prior. In the
limit, the acceptance ratio reduces to that of
Algorithm~\ref{alg:fixing-bnu}. One could choose $\alpha$ based on
preliminary runs, but such tuning would be burdensome.

For likelihood terms that factorize, $\lh(\gv)\te\prod_i \lh_i(f_i)$,
we can measure how much the likelihood restricts each variable
individually:
\begin{equation}
    P(f_i\g \lh_i,\theta) \propto \lh_i(f_i)\;\N(f_i; 0, (\Sprior)_{ii}).
    \label{eqn:site_posterior}
\end{equation}
A Gaussian can be fitted by moment matching or a Laplace approximation
(matching second derivatives at the mode). Such fits, or close
approximations, are often possible analytically and can always be
performed numerically as the distribution is only one-dimensional.
Given a Gaussian fit to the site-posterior~\eqref{eqn:site_posterior}
with variance~$v_i$, we can set the auxiliary noise to a level that
would result in the same posterior variance at that site alone:
${(\Snoise)_{ii}\te(v_i^{-1} \tm {(\Sprior)_{ii}}^{-1})^{-1}}$.
(Any negative $(\Snoise)_{ii}$ must be thresholded.)
The moment matching procedure is a grossly
simplified first step of ``assumed density filtering'' or
``expectation propagation'' \citep{minka2001b}, which are too
expensive for our use in the inner-loop of a Markov chain.

\vspace*{-0.15cm}
\section{Related work}
\vspace*{-0.25cm}

We have discussed samplers that jointly update strongly-coupled latent
variables and hyperparameters. The hyperparameters can move further in
joint moves than their narrow conditional posteriors (e.g.,
Figure~\ref{fig:postcondf}) would allow. A generic way of jointly
sampling real-valued variables is Hamiltonian/Hybrid Monte Carlo (HMC)
\citep{duane1987,neal2011}. However, this method is cumbersome to
implement and tune, and using HMC to jointly update latent variables
and hyperparameters in hierarchical models does not itself seem to improve
sampling \citep{choo2000}.

\Citet{christensen2006} have also proposed a robust representation for
sampling in latent Gaussian models. They use an approximation to the
target posterior distribution to construct a reparameterization where
the unknown variables are close to independent.  The approximation
replaces the likelihood with a Gaussian form proportional to $\N(\gv;
\hat{\gv}, \Lambda(\hat{\gv}))$:
\begin{equation}
    \textstyle
    \hat{\gv} = \argmax_\gv \lh(\gv),\qquad
    \Lambda_{ij}(\hat{\gv}) = \left.\frac{\partial^2 \log\lh(\gv)}{\partial f_i\, \partial f_j}\right|_{\hat{\gv}},
    \label{eqn:taylor}
\end{equation}
where $\Lambda$ is often diagonal, or it was suggested one would only
take the diagonal part. This Taylor approximation looks like a Laplace
approximation, except that the likelihood function is not a
probability density in~$\gv$.
This likelihood fit results in an approximate Gaussian posterior
$\N(\gv; \bmm_{\theta\kern-1pt,\gvv=\hat{\gv}}, \Sppost)$ as found in%
~\eqref{eqn:pseudopost}, with
noise $\Snoise\te\Lambda(\hat{\gv})^{-1}$ and
data $\gvv\te\hat{\gv}$.

Thinking of the current latent variables as a draw from this
approximate posterior, ${\bomega\tsim\N(0,\eye)},\hspace{1ex} \gv \te
\Lppost\bomega \tp \bmm_{\theta\kern-1pt,\hat{\gv}}$, suggests using the
reparameterization ${\bomega = \Lppost^{-1}(\gv - \bmm_{\theta\kern-1pt,\hat{\gv}})}$%
.
We can then fix the new variables and update the hyperparameters under
\begin{equation}
    P(\theta \g \bomega, \D) \propto
    \lh(\gv(\bomega,\theta))\,\N(\gv(\bomega,\theta); 0,\Sprior)\,\ph(\theta) \left| \Lppost \right|.
    \label{eqn:fixed_score}
\end{equation}
When the likelihood is Gaussian, the reparameterized variables
$\bomega$ are independent of each other and the hyperparameters. The
hope is that approximating non-Gaussian likelihoods will result in
nearly-independent parameterizations on which Markov chains will mix
rapidly.

Taylor expanding some common log-likelihoods around the maximum is not
well defined, for example approximating probit or logistic likelihoods
for binary classification, or Poisson observations with zero counts.
These Taylor expansions could be seen as giving flat or undefined
Gaussian approximations that do not reweight the prior. When all of
the likelihood terms are flat the reparameterization approach reduces
to that of Section~\ref{sec:whiten}. The alternative $\Snoise$
auxiliary covariances that we have proposed could be used instead.

The surrogate data samplers of Section~\ref{sec:surrogate} can also be viewed as using
reparameterizations, by
treating $\etab\te\Lppost^{-1}(\gv-\mppost)$ as an arbitrary
random reparameterization for making proposals. A proposal density
$q(\etab',\theta';\etab,\theta)$ in the reparameterized space must be
multiplied by the Jacobian $|\Lppostp^{-1}|$ to give a proposal
density in the original parameterization. The probability of proposing
the reparameterization must also be included in the
Metropolis--Hastings acceptance probability:
\begin{equation}
    \textstyle \min\left(1,\; \frac%
    {P(\theta',\gv' \g \D)\cdot P(\gvv\g\gv',\Snoisep) \cdot q(\theta;\theta')\,|\Lppost^{-1}|}%
    {P(\theta,\gv \g \D)\cdot P(\gvv\g\gv,\Snoise) \cdot q(\theta';\theta)\,|\Lppostp^{-1}|}
    \right).
    \label{eqn:reparam_view_accept}
\end{equation}
A few lines of linear algebra confirms that, as it must do, the same
acceptance ratio results as before.
Alternatively, substituting \eqref{eqn:joint_posterior} into
\eqref{eqn:reparam_view_accept} shows that the acceptance probability
is very similar to that obtained by applying Metropolis--Hastings to
\eqref{eqn:fixed_score} as proposed by \citet{christensen2006}.
The differences are that the new latent variables~$\gv'$ are
computed using different pseudo-posterior means and the surrogate
data method has an extra term for the random, rather than fixed,
choice of reparameterization.

The surrogate data sampler is easier to implement than the previous
reparameterization work because the surrogate posterior is centred
around the current latent variables. This means that 1)~no point
estimate, such as the maximum likelihood~$\hat{\gv}$, is required.
2)~picking the noise covariance~$\Snoise$ poorly may still produce a
workable method, whereas a fixed reparameterized can work badly if the
true posterior distribution is in the tails of the Gaussian
approximation. \Citet{christensen2006} pointed out that centering the
approximate Gaussian likelihood in their reparameterization around the
current state is tempting, but that computing the Jacobian of the
transformation is then intractable. By construction, the surrogate
data model centers the reparameterization near to the current state.

\section{Experiments}

\begin{figure}[t!]
\centering%
\includegraphics[width=\textwidth]{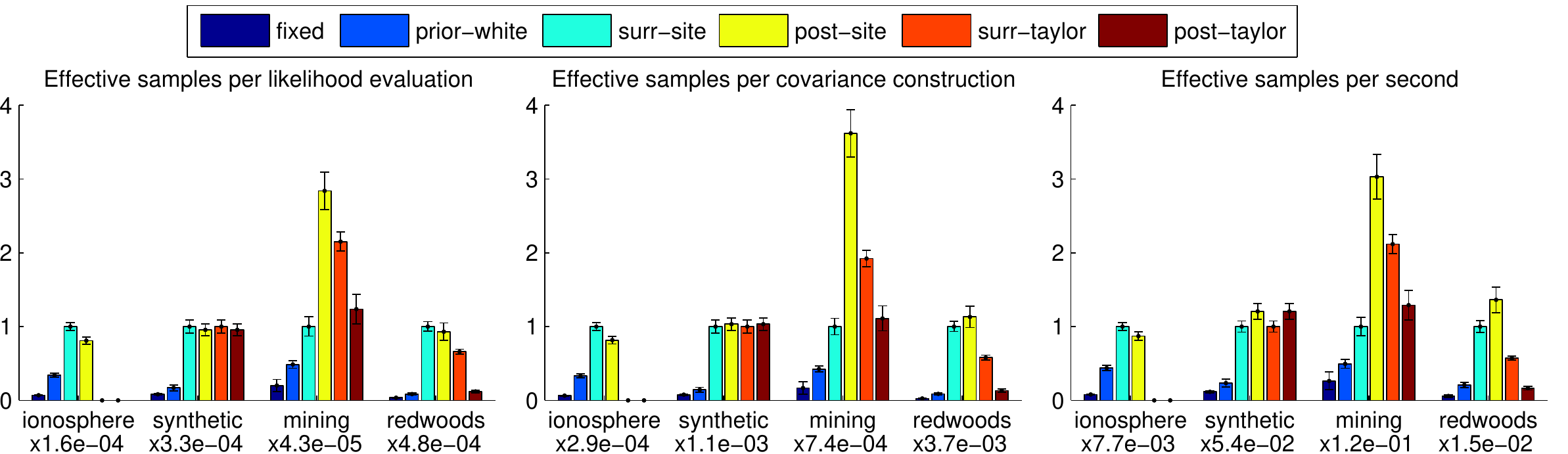}%
\caption{\small The results of experimental comparisons of six MCMC
  methods for GP hyperparameter inference on four data sets.  Each
  figure shows four groups of bars (one for each experiment) and the
  vertical axis shows the effective number of samples of the complete
  data likelihood per unit cost.  The costs are \textit{per likelihood
    evaluation} (left), \textit{per covariance construction}
  (center), and \textit{per second} (right).
  Means and standard errors for 10 runs are shown.
  Each group of bars has been rescaled for readability: the number
  beneath each group gives the effective samples for the
  \texttt{surr-site}
  method, which always has bars of height~1.  Bars are missing where
  methods are inapplicable (see text).}%
\label{fig:results}%
\vspace*{-0.1cm}
\end{figure}

We empirically compare the performance of the various approaches to GP
hyperparameter sampling on four data sets: one regression, one
classification, and two Cox process inference problems. Further
details are in the rest of this section, with full code as
supplementary material. The results are summarized in
Figure~\ref{fig:results} followed by a discussion section.

In each of the experimental configurations, we ran ten independent
chains with different random seeds, burning in for 1000 iterations and
sampling for 5000 iterations.  We quantify the mixing of the chain by
estimating the effective number of samples of the complete data
likelihood trace using R-CODA~\cite{cowles2006}, and compare that with
three cost metrics: the number of hyperparameter settings considered
(each requiring a small number of covariance decompositions
with~$O(n^3)$ time complexity), the number of likelihood evaluations,
and the total elapsed time on a single core of an Intel Xeon
3GHz~CPU\@.

The experiments are designed to test the mixing of
hyperparameters~$\theta$ while sampling from the joint
posterior~\eqref{eqn:joint_posterior}. All of the discussed approaches
except Algorithm~\ref{alg:fixing-gv} update the latent variables~$\gv$
as a side-effect. However, further transition operators for the latent
variables for fixed hyperparameters are required. In
Algorithm~\ref{alg:fixing-bnu} the ``whitened'' variables~$\bnu$
remain fixed; the latent variables and hyperparameters are constrained
to satisfy $\gv \te \Lprior \bnu$. The surrogate data samplers are
ergodic: the full joint posterior distribution will eventually be
explored. However, each update changes the hyperparameters and
requires expensive computations involving covariances. After computing
the covariances for one set of hyperparameters, it makes sense to
apply several cheap updates to the latent variables.
For every method we applied ten updates of elliptical slice sampling
\citep{murray2010} to the latent variables~$\gv$ between each
hyperparameter update. One could also consider applying elliptical
slice sampling to a reparameterized representation, for simplicity of
comparison we do not.
Independently of our work \citet{titsias2010} has used surrogate
data like reparameterizations to update latent variables
for fixed hyperparameters.

\paragraph{Methods} We implemented
six
methods for updating Gaussian
covariance hyperparameters. Each method used the same slice sampler,
as in Algorithm~\ref{alg:surr-slice}, applied to the following model
representations.
\textbf{fixed:}~fixing the latent function~$\gv$ \cite{neal1999a}.
\textbf{prior-white:}~whitening with the prior.
\textbf{surr-site:}~using surrogate data with the noise level set to
match the site posterior~\eqref{eqn:site_posterior}. We used Laplace
approximations for the Poisson likelihood. For classification problems
we used moment matching, because Laplace approximations do not work
well \citep{kuss2005}.
\textbf{surr-taylor:}~using surrogate data with noise variance set via
Taylor expansion of the log-likelihood~\eqref{eqn:taylor}. Infinite variances were
truncated to a large value.
\textbf{post-taylor} and \textbf{post-site:} as for the \texttt{surr-} methods
but a fixed reparameterization based on a posterior approximation~\eqref{eqn:fixed_score}.

\paragraph{Binary Classification (Ionosphere)}
We evaluated four different methods for performing binary GP
classification:
\texttt{fixed},
\texttt{prior-white},
\texttt{surr-site} and
\texttt{post-site}.
We applied these methods to the Ionosphere dataset
\citep{sigillito1989}, using 200 training data and 34 dimensions.  We
used a logistic likelihood with zero-mean prior, inferring
lengthscales as well as signal variance. The \texttt{-taylor} methods
reduce to other methods or don't apply because the maximum of the
log-likelihood is at plus or minus infinity.

\paragraph{Gaussian Regression (Synthetic)}
When the observations have Gaussian noise the \texttt{post-taylor} reparameterization of
\citet{christensen2006} makes the hyperparameters and latent variables
exactly independent. The random centering of the surrogate data model
will
be
less effective.
We used a Gaussian regression
problem to assess how much worse the surrogate data method is compared
to an ideal reparameterization. The synthetic data set had 200 input
points in 10-D drawn uniformly within a unit hypercube. The GP had
zero mean, unit signal variance and its ten lengthscales
in~\eqref{eqn:se_kernel} drawn from~$\U(0,\surd{10})$. Observation
noise had variance~0.09.  We applied the \texttt{fixed},
\texttt{prior-white}, \texttt{surr-site}/\texttt{surr-taylor}, and
\texttt{post-site}/\texttt{post-taylor} methods. For Gaussian likelihoods the
\texttt{-site} and
\texttt{-taylor} methods coincide: the auxiliary noise matches the
observation noise~($\Snoise=0.09\,\eye$).

\paragraph{Cox process inference}
We tested all six methods on an
inhomogeneous Poisson process with a Gaussian process prior for the
log-rate.
We sampled the hyperparameters in~\eqref{eqn:se_kernel} and a mean
offset to the log-rate. The model was applied to two point process
datasets: 1)~a record of mining disasters \citep{jarrett1979} with
191~events in 112~bins of 365~days. 2)~195 redwood tree locations in a
region scaled to the unit square \citep{ripley1977} split into
$25\ttimes25\te625$ bins. The results for the mining problem were
initially highly variable. As the mining experiments were also the
quickest we re-ran each chain for 20,000 iterations.

\section{Discussion}
\vspace*{-0.1cm}

On the Ionosphere
classification problem
both of the \texttt{-site} methods worked
much better than the two baselines. We slightly prefer
\texttt{surr-site} as it involves less problem-specific derivations
than \texttt{post-site}.

On the synthetic test the \texttt{post-} and \texttt{surr-} methods
perform very similarly. We had expected the existing \texttt{post-}
method to have an advantage of perhaps up to 2--3$\times$, but that was not
realized on this particular dataset. The \texttt{post-} methods had a
slight time advantage, but this is down to implementation details and
is not notable.

On the mining problem the Poisson likelihoods are often close to
Gaussian, so the existing \texttt{post-taylor} approximation works
well, as do all of our new proposed methods.  The Gaussian approximations to the Poisson
likelihood fit most poorly to sites with zero
counts. The redwood dataset discretizes two-dimensional space, leading
to a large number of bins.  The majority of these bins have zero
counts, many more than the mining dataset. Taylor expanding the
likelihood gives no likelihood contribution for bins with zero counts,
so it is unsurprising that \texttt{post-taylor} performs similarly to
\texttt{prior-white}. While \texttt{surr-taylor} works better, the
best results here come from using approximations to the
site-posterior~\eqref{eqn:site_posterior}. For unreasonably fine
discretizations the results can be different again: the
\texttt{site-} reparameterizations do not \emph{always} work well.

Our empirical investigation used slice sampling because it is easy to
implement and use. However, all of the representations we discuss
could be combined with any other MCMC method, such as
\citep{girolami2011} recently used for Cox processes. The new
surrogate data and \texttt{post-site} representations offer
state-of-the-art performance and are the first such advanced methods
to be applicable to Gaussian process classification.

An important message from our results is that fixing the
latent variables and updating hyperparameters according to the
conditional posterior\,---\,as commonly used by GP
practitioners%
\,---\,can work exceedingly poorly.  Even the simple
reparameterization of ``whitening the prior'' discussed in
Section~\ref{sec:whiten} works much better on problems where
smoothness is important in the posterior.  Even if site approximations
are difficult and the more advanced methods presented are
inapplicable, the simple whitening reparameterization should be given
serious consideration when performing MCMC inference of
hyperparameters.

\ifnipsfinal
\subsubsection*{Acknowledgements}
\vspace*{-0.1cm}

We thank an anonymous reviewer for useful comments.
This work was supported in part by the IST Programme of the European
Community, under the PASCAL2 Network of Excellence, IST-2007-216886.
This publication only reflects the authors' views.
RPA is a junior fellow of the Canadian Institute for Advanced Research.

\fi

\let\origurl\url
\renewcommand{\url}[1]{\penalty10000 \hskip.5em
    plus\linewidth \interlinepenalty10000\penalty200
    \hskip-.17em plus-\linewidth minus.11em \origurl{#1}}

\bibliographystyle{unsrtnat}
\bibliography{bibs}

\end{document}